\documentclass{emulateapj}

\usepackage{amsmath}
\usepackage{rotating}

\newcommand{\chandra}{{\em Chandra}}
\newcommand{\suzaku}{{\em Suzaku}}
\newcommand{\fermi}{{\em Fermi}}
\newcommand{\gray}{{$\gamma$-ray}}
                                   
\shortauthors{J. Vink et al.}

\begin{document}

\title{
The radiative X-ray and gamma-ray efficiencies of rotation powered pulsars}

\author{Jacco Vink$^1$, Aya Bamba$^{2,3}$, Ryo Yamazaki$^4$}
\affil{$^1$Astronomical Institute, Utrecht University, P.O. Box 80000, 
3508TA Utrecht, The Netherlands\\
$^{2}$School of Cosmic Physics, Dublin Institute for Advanced Studies 31 Fitzwilliam Place, Dublin 2, Republic of Ireland\\
$^3$ ISAS/JAXA Department of High Energy Astrophysics 3-1-1 Yoshinodai, Chuo-ku,
Sagamihara, Kanagawa 252-5210, Japan\\
$^4$ Department of Physics and Mathematics, Aoyama Gakuin University, 5-10-1 Fuchinobe, Chuo-ku, Sagamihara, Kanagawa, 252-5258, Japan
}

\email{j.vink@astro-uu.nl}

\begin{abstract}
We present a statistical analysis of the X-ray luminosity of rotation powered
 pulsars and their surrounding nebulae
using the sample of \citet{kargaltsev08} and we complement this with an analysis 
of the \gray-emission of \fermi\ detected pulsars.
We report a strong trend in the efficiency with which spin-down
power is converted to X-ray and \gray\ emission with characteristic
age: 
young pulsars and their surrounding nebulae are efficient
X-ray emitters, whereas in contrast 
old pulsars are efficient \gray\ emitters.
We divided the X-ray sample in a young ($\tau_c < 1.7\times 10^4$~yr)
and old sample and used 
linear regression to search for correlations 
between the  logarithm of the
X-ray and \gray\ luminosities and the logarithms of
the periods and period derivatives. 
The X-ray emission from young pulsars and their nebulae
are both consistent with $L_X \propto \dot{P}^3/P^6$.
For old pulsars and their nebulae the X-ray luminosity is consistent
with a more or less constant efficiency 
$\eta \equiv L_X/\dot{E}_{rot} \approx 8\times 10^{-5}$.
For the \gray\ luminosity we confirm that 
$L_\gamma \propto \sqrt{\dot{E}_{rot}}$. 

We discuss these findings in the context of pair production inside pulsar 
magnetospheres and the striped wind model. 
We suggest that the striped wind model may explain the similarity between the 
X-ray properties of the pulsar wind nebulae and the pulsars themselves, 
which according to the striped wind model may both find their origin outside
the light cylinder, in the pulsar wind zone.
\end{abstract}


\section{Introduction}

Despite more than four decades of research, the emission from
pulsars and their surrounding nebulae is still poorly understood.
Two important problems are the origin of pulsar \gray\ emission
and,
for  pulsar wind nebulae (PWNe),
the so-called sigma-problem \citep[see the reviews by][]{gaensler06,arons09,kirk09}. 
The latter problem
derives its name from the sigma-parameter, 
$\sigma \equiv B^2/4\pi \Gamma  m_{\rm e}c^2 n$, 
which is the  ratio of the energy density due to Poynting flux over the
particle energy density, with $B$ the local magnetic field strength,
$\Gamma$ the Lorentz factor of the pulsar wind,
and $n$ the particle number density.
According to theoretical models, most of the pulsar's rotational
energy  loss rate (or spin-down power) $\dot{E}_{rot}$ 
is due to Poynting flux, i.e. the pulsar wind should
have $\sigma >> 1$,
but observations of PWNe indicate that most of the energy that has been lost 
is actually contained by the relativistic electrons/positrons.
Somehow, the high $\sigma$ flow is converted into kinetic energy
somewhere between the pulsar's magnetosphere and the
wind termination shock, which converts the radial pulsar wind into
an isotropic, relativistic particle
distribution \citep{arons09,kirk09}. 

A third, and perhaps related, problem is
the high wind multiplicity factor. The combination
of a high magnetic and rapid rotation results in 
a strong electric potential in the magnetosphere. This potential 
will be neutralized by
charged particles that are stripped from the surface of the neutron star
\citep{goldreich69}. The associated charged particle density is
\begin{equation}
n_{GJ}= 7\times 10^{-2} B/P~{\rm cm}^{-3},
\end{equation} 
with $P$ the pulsar's period and $B$ the local magnetic field.
A fraction of these particles will escape through open field lines,
resulting in a particle flux  
\begin{equation}
\dot{N}_{GJ} =2.7\times 10^{30} P^{-2}B_{12}~{\rm s}^{-1},
\end{equation}
with $B_{12}$ the dipole surface  magnetic field in units of $10^{12}$~G.
However, X-ray \citep[e.g.][]{gaensler02} and TeV \citep{dejager07} observations
indicate that the number of relativistic electrons contained by PWNe turns out
to be orders of magnitude larger than $\dot{N}_{GJ}$, i.e. 
$\dot{N} = \kappa \dot{N}_{GJ}$, with the multiplicity
factor being $\kappa > 500$ for a young pulsar like
B1509-58 \citep{dejager07}. The origin of the additional plasma is likely
electron/positron pair production in the magnetosphere.
The pair production occurs in the presence of the high
magnetic fields inside the magnetosphere, and requires the presence of
high energy
photons that are either the result of
curvature radiation or inverse Compton scattering. The
electrons that cause the emission are
accelerated due to the extremely large voltage drop across the open field lines
\citep{hibschman01,harding02}. For the inverse Compton scattering
seed photons are necessary that are emitted by the hot polar caps of the
pulsar, heated due to the bombardment by accelerated particles,
or due to the cooling of the young neutron star.

Despite the many unsolved problems, pulsar research has thrived over the last decade thanks
to many advances in observational techniques and numerical simulations.
In particular high energy observations have contributed to a wealth
of new information on pulsars and PWNe; from high spatial resolution
X-ray images with \chandra, revealing torii and jets 
\citep[e.g.][]{hester02,helfand01,bamba10},
to a rapid progress in the field of TeV astronomy \citep[][for a review]{hinton09}, 
which have revealed
an unexpectedly large population of very extended PWNe 
\citep[e.g.][]{dejager07,mattana09}. This rapid growth in observational
data has recently been augmented by the GeV \gray\ observatory \fermi,
which has greatly increased the sample of \gray-emitting
pulsars \citep{abdo10}.

Here we present a statistical analysis of two samples of rotation powered
pulsars. One is those 
of X-ray pulsars compiled by \citet{kargaltsev08}, the other
the aforementioned sample of \fermi-detected pulsars. Our analysis
concentrates on what determines the radiative efficiency of pulsars and their
PWNe.
We report a surprisingly strong correlation between
the X-ray luminosity of pulsars and their PWNe, which inversely
correlates with characteristic age, at least for young pulsars.
In contrast, the \gray\ emission correlates positively with 
characteristic pulsar age, as already noted by \citet{abdo10}.

\begin{figure*}[t]
\centerline{
\includegraphics[angle=-90,width=0.75\textwidth]{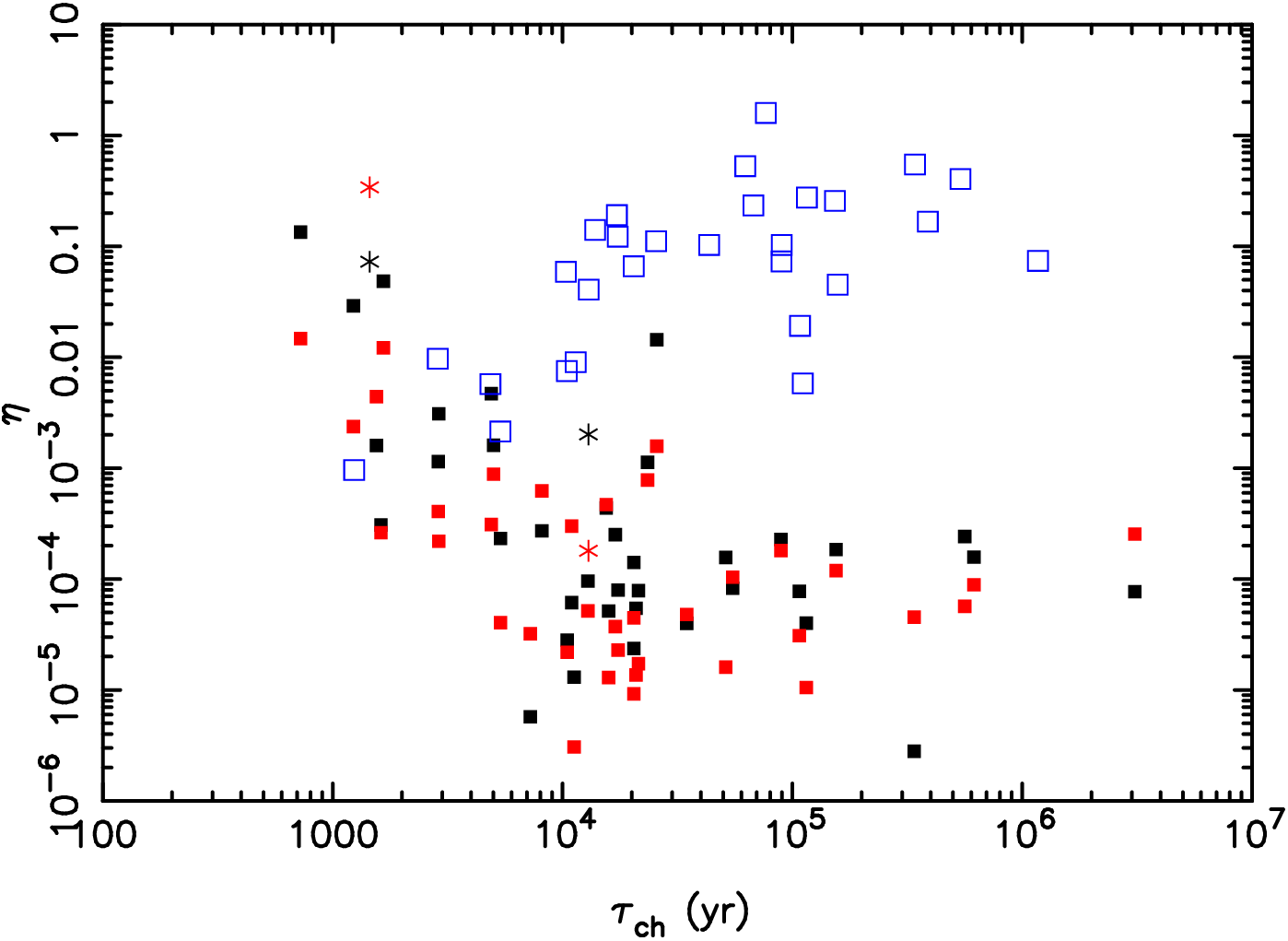}
}
\caption{
The X-ray radiation efficiency $\eta$ of PWNe (solid black squares) and
pulsars (solid red squares) as a function of characteristic age $\tau_c$.
The data has been taken from \citet{kargaltsev08} and is based on
the X-ray luminosity in the 0.5-8 keV band as measured by \chandra.
The open blue squares indicate the 
\gray\ efficiencies in the 100 MeV to 100 GeV band
based on \fermi\ data \citep{abdo10}.
The star-like symbols indicate two recently discovered pulsars (black) and
PWNe (red) which
were not part of the \citet{kargaltsev08} sample: 
the AXP 1E1547.0- 5408 \citep[$\tau_{ch}=1447$~yr][]{vink09,camilo07},
and PSR J14003-6326 \citep[$\tau_{ch}=23.7$~kyr][]{renaud10}. 
For these two pulsars/PWN the reported unabsorbed
fluxes were converted to a 0.5-8 keV band luminosity using the distances
quoted in the papers.
\label{fig:eta}
}
\end{figure*}

\section{A statistical analysis}
It is well known that the non-thermal X-ray luminosities of
pulsars and PWNe are strongly correlated with the spin-down luminosity 
$\dot{E}_{rot}$ of the pulsar 
\citep{seward88,verbunt96,becker97,possenti02,cheng04,kargaltsev08}.
The efficiency with which the spin-down luminosity is converted into
X-ray emission is usually indicated by the symbol 
\begin{equation}
\eta \equiv L_X/\dot{E}_{rot},
\label{eq:eta}
\end{equation}
with $\eta$ in the range of  $\eta \sim 10^{-6}- 1$.
An important question is what determines this efficiency for both
\gray\ and X-ray emission. For
the PWNe $\eta$ may provide information on how well the spin-down
luminosity is converted into relativistic particles, i.e. it is related to
the sigma-problem.

For the statistical analysis described here we use the X-ray properties
of pulsars and their nebulae as determined by \citet{kargaltsev08}.
This sample is based on \chandra\ observations. 
The \chandra-ACIS instrument that was used has a poor timing resolution, so
all pulsar luminosities are a combination of pulsed and unpulsed 
emission. However, given the high spatial
resolution of \chandra, the X-ray flux from the pulsar could be accurately
separated from the X-ray flux from the PWN. The X-ray luminosities were derived
from the 0.5-8 keV fluxes, corrected for interstellar absorption and using the
distance estimated listed in the paper. \citet{kargaltsev08} list only
the non-thermal X-ray luminosities, 
so the contribution of thermal X-ray emission, if present, was ignored.
The uncertainties in the distance are
the largest source of error in the luminosities. In general,
the distance estimates may have errors of order $\sim 2$, resulting
in luminosity errors of order 4.
Of course distance errors affect the
pulsar and PWNe luminosities in the same way.
A source of error for the PWNe luminosities may be underestimated flux
contributions from
 low surface brightness emission at large radii. Indeed,  
\citet{bamba10b} recently reported the detection of low X-ray surface brightness
structures around several PWNe by \suzaku. The fraction of the luminosity
in these low surface brightness is, however, not more than a factor
of 2 in luminosity. So all together, the logarithm of the luminosities may 
contain errors of the order of one decade.

We included in our study a statistical analysis of the \gray\ luminosities,
based on the pulsed \gray\ emission of pulsar reported by \citet{abdo10}.
We made sure that the \gray\ luminosities of pulsars that were common to both
the X-ray and \gray\ samples were based on the same
distance estimates, namely those adopted by \citet{kargaltsev08}. 
For both the X-ray and \gray\ samples we excluded 
millisecond pulsars, and omitted
pulsars without good distance estimates.

In figure~\ref{fig:eta} we show the correlation of the X-ray luminosity
efficiency $\eta$ of pulsars and PWNe versus the characteristic spin-down
age $\tau_c= P/2\dot{P}$. {
We also included data points for two recently discovered
PWNe; those surrounding the high energy pulsar J14003-6326 \citep{renaud10} 
and AXP 1E1547.0- 5408 \citep{vink09}. We did not include these in our
statistical analysis, although their properties are consistent with the
general trends we report below.

Figure~\ref{fig:eta} also shows the \gray\ efficiency, 
as obtained from the pulsed
\gray\ luminosities determined by \citet{abdo10}.
This figure reveals the trend that young pulsars appear to
have higher X-ray efficiencies than old pulsars. Moreover, the behavior
is similar for the luminosities of the pulsars and the PWNe.

\begin{figure}[b]
\centerline{
\includegraphics[angle=-90,width=\columnwidth]{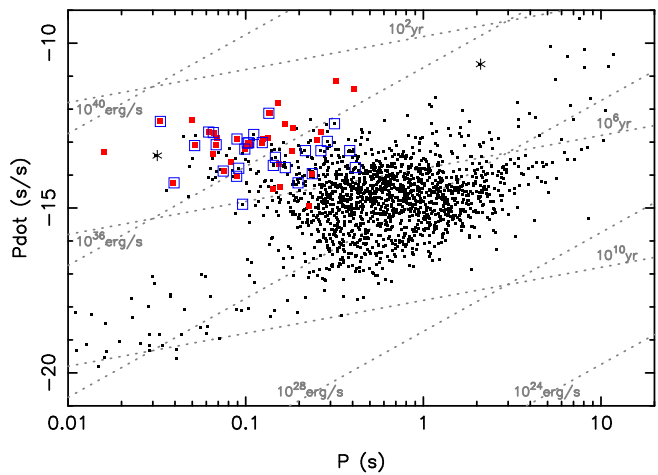}
}
\caption{
The period and period derivative of the pulsars in the
X-ray \citep{kargaltsev08} and \gray\ \citep{abdo10} samples.
The symbols match those in Fig.~\ref{fig:eta}. Star-like symbols represent
 AXP 1E1547.0- 5408 (upper right) and PSR J14003-6326.
The small dots represent pulsars drawn from the ATNF catalogue
\citep{manchester05}.
\label{fig:ppdot}
}
\end{figure}

Figure~\ref{fig:ppdot} shows the timing properties of the pulsars
in the two samples used for our statistical analysis in the 
$P-\dot{P}$ diagram. It shows that on average the detected \gray\ pulsars
seem to be somewhat older than the X-ray detected pulsars/PWNe, consistent
with the trend that older pulsars are less efficient in emitting X-rays
and more efficient in emitting \gray s.
Note, however, that
an interpretation of this diagram is far from  straightforward, 
because the detectability of
a pulsar also depends on its distance and, for X-rays, 
on the interstellar absorption column.}

A trend of decreasing X-ray luminosity with $\tau_c$ was reported
before by \citet{becker97,possenti02}, but figure~\ref{fig:eta}
reveals it to be due to a trend in the efficiency, not in the overall
spin-down power. 
For pulsars with $\tau_c \gtrsim 2\times 10^4$~yr
the efficiency appears to be more or less constant.
In contrast, young pulsars are not so efficient in producing \gray\ emission,
as already noted by \citet{abdo10}. 
The spread in the data points for
a given characteristic is of order of 1-2 decades. 
As discussed above, the uncertainties 
in the luminosity can explain about 1 decade of this spread, but
intrinsic variations, and for the \gray\  emission, beaming properties
are expected to contribute to the spread as well.

In order to investigate the correlations between $\eta$ and $\tau_c$ 
further, one has to
avoid $\eta$ as an independent variable, since
both $\eta$ and $\tau_c$ are derived from combinations of the pulsar's period
$P$ and period derivative $\dot{P}$; the spin-down luminosity is given by
\begin{equation}
\dot{E}_{rot} = I \dot{\Omega} \Omega  = 4\pi^2 I\frac{\dot{P}}{P^3},
\end{equation}
with $I\approx 1.4\times 10^{45}$~g\,cm$^2$ the neutron star's moment of inertia.
For that reason we base our regression analysis 
on the logarithm\footnote{
We use $\log$ to denote the 10-based logarithm.} 
of the X-ray/\gray\ 
luminosity, as the quantity to be explained, 
and the independent variables $\log P$, $\log \dot{P}$ as the
principle input  variables for the model \citep[c.f.][]{possenti02}.

As explained above the errors are dominated by systematic errors,
mostly due to difficulties in estimating distances.
For that reason we used the unweighted least square method,
which means that 
the errors in the best-fit parameters are based on the variance of
the residuals.
The disadvantage is that we do not have an intrinsic goodness of
fit statistic. However, 
we can compare two hypotheses using the F-test statistic, defined as:
\begin{equation}
F = \frac{
\sum_i  (\log L_i -\log \tilde{L}_{1i})^2/(m_2-m_1)}{\sum_i  (\log L_i - \log \tilde{L}_{2i})^2/(n-m_2)},
\end{equation}
here $L_i$ denotes the observed (\gray/X-ray) luminosity and
we use $\tilde{L}_{1i}$ to indicate the expected value of $L_i$ based on the best-fit parameters,
and the subscript $1$ and $2$ to indicate two different best-fit models, $m$ is the number of degrees of
freedom and $n$ is the sample size. The probability that the
improvement in the sum of squared residuals is by pure chance is given by the $F(m_2-m_1,n-m_2)$-distribution.

From the above it is clear that we fitted the linear relation:
\begin{equation}
\log L =  a + b \log P + c \log \dot{P}.
\label{eq:fullfit}
\end{equation}
We compare this relation, using the F-test, with a functional form $\log L - \log  \dot{E}_{rot} = a$.
This is essentially assuming that $\eta$ (equation~\ref{eq:eta}) is constant. 
In addition we fit the relation
\begin{equation}
\log L =  a + b \log \dot{E}_{rot},\label{eq:edotfit}
\end{equation}
which corresponds to $L \propto \dot{E}_{rot}^b \propto \dot{P}^{b}/P^{3b}$. 
This is a functional form that
has been used before \citep[e.g.][]{kargaltsev08}, 
and that seems to work particularly well for
the \gray\ emission from pulsars \citep{arons96,abdo10}.

Since figure~\ref{fig:eta} suggests that young pulsars are more efficient
X-ray emitters than old pulsars, we divided the X-ray sample in an old and a young population. As the precise 
characteristic age that should be chosen is a bit arbitrary, but lies somewhere
between $10^4$~yr and $5\times 10^4$~yr, we decided to
divide the sample in two more or less equally sized samples. This put
the cut at  $\tau_c=1.7\times 10^4$~yr.

\begin{figure}[b]
\centerline{
\includegraphics[angle=-90,width=\columnwidth]{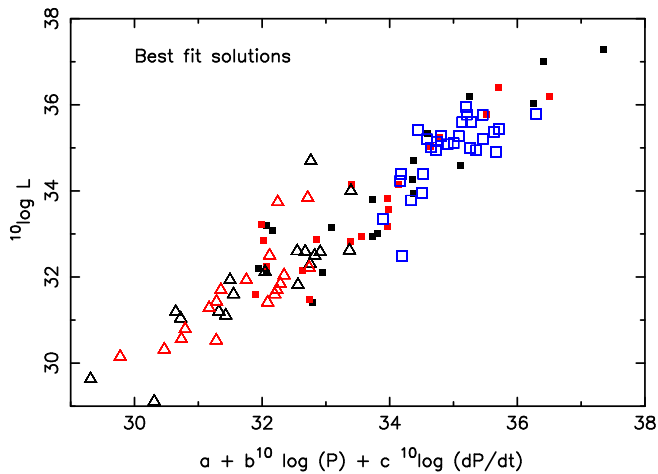}
}
\caption{
The data sample plotted as $\log L_X,\log L_\gamma$ versus the best-fit 
models using equation~\ref{eq:fullfit}. The symbols indicate:
black squares: luminosity of  young PWNe; red squares: 
luminosity of young pulsars;
black open triangles: luminosity of old PWNe; red open triangles: luminosity of olds pulsars; blue open squares: \gray\ pulsars.
\label{fig:bestfit}
}
\end{figure}

The result of the regression analysis is listed in Table~\ref{tab:fits},
whereas figure~\ref{fig:bestfit} shows the results of applying the best-fit solutions
to the different samples.
The best-fit relations for young pulsars and PWNe indicate that the X-ray luminosity
does not scale with $\dot{E}_{rot}$, 
since in that case we would expect for equation (\ref{eq:fullfit}) 
$b \approx -3$ and $c \approx 1$,
whereas we find $b = -6.2\pm 0.7$ and $c = 3.3\pm0.4$ and 
$b = -5.2\pm 0.7$ and $c = 2.8\pm 0.4$  for PWNe and pulsars,
respectively. 
This conclusion is based on both the best-fit parameters and their inferred
errors, as on the F-test.  The F-test, in fact, indicates that  equation (\ref{eq:fullfit}) 
provides a better fit than equation (\ref{eq:eta}) at the 99.99\% confidence level.
Interestingly, the
values for $b$ and $c$ for pulsars and PWNe are consistent with each other.

For older pulsars the F-values indicate that the 
X-ray emission efficiency is just as well described
by equation (\ref{eq:eta}) with $\eta \approx 8\times 10^{-5}$, 
as by a linear dependency on $\log P$ and $\log \dot{P}$.
Also this result is true for both the pulsar X-ray luminosity as
for the PWN luminosity.

The least square fits for the \fermi-sample shows that the 
relation $L_\gamma \propto \sqrt{\dot{E}_{rot}}$ is indeed a good description
of the data.
For equation~(\ref{eq:edotfit}) we find $b=0.51\pm 0.10$, close
to the expected 1/2, and for equation~(\ref{eq:fullfit}) we find
$b=-1.6\pm 0.4$ and $c=0.49\pm 0.18$, whereas  a scaling with $\sqrt{\dot{E}_{rot}}$ implies
$b=-3/2$ and $c=1/2$, in excellent agreement.
The F-tests confirm this, and indicate that a constant value for 
$\eta_\gamma$ can be rejected at the 99.9996\% confidence level.

Finally, one may wonder whether the combined \gray\ and X-ray luminosity 
is more closely correlated with $\dot{E}_{rot}$, than the \gray\ and X-ray luminosities
individually. Unfortunately, this is hard to determine from the present data
sets as there are only 15 pulsars that the two samples have in common, with most of them being
older pulsars. In fact,  the combined X-ray/\gray\ luminosity of all pulsars
is dominated by the \gray\ emission, with the exception of the Crab pulsar.
It is, therefore, not surprising that the best-fit relation between $L_{X,\gamma}$
indicates $L_{X+\gamma} \propto \dot{E}^{1/2}_{rot}$, just like for \gray\ 
luminosity. 
However, as the number of \gray-detected pulsars will grow in the near future,
it may be good to investigate the total radiative output from rotation powered pulsars
in more detail, especially around $\tau_c \approx 10^4$~yr, where the radiative
output changes from X-ray to \gray\ domination.

\section{Discussion}
We presented a statistical study of the X-ray and \gray\ properties
of rotation powered pulsars, with the aim of finding what trends
underly the efficiency with which spin-down luminosity 
is converted into high energy radiation.
We started our analysis by showing that young pulsars
are efficient X-ray emitters, but poor \gray\ emitters,
whereas it is the other way around for old pulsars.
A trend of low X-ray luminosity as a function of characteristic
age was reported by \citet{becker97,possenti02}, but it is shown
here that it is related to the efficiency with which spin-down
luminosity is converted to X-ray emission, and not just due to an overall
decline in spin-down luminosity.

In order to be as general as possible in finding trends in 
X-ray/\gray luminosities our main results are based
on a regression analysis of X-ray luminosity of the pulsars/PWNe
\citep{kargaltsev08}
and \gray\ luminosity of pulsars \citep{abdo10} versus
the pulse period and its derivative. 
This is a very generic method, which includes the possibility that
the luminosity depends solely on spin-down luminosity.
Thus encompassing other possible
dependencies, such as on spin-down luminosity.
Given the trend noted in figure~\ref{fig:eta} we divided the X-ray
sample in two equally sized samples of young ($\tau_c < 1.7\times 10^4$~yr)
and old pulsars.

Our statistical analysis produced two new findings: 
1) the X-ray luminosity of pulsars and
their surrounding PWNe appear closely correlated, with $L_{X,psr/pwne}\propto \dot{P}^3/P^6$
for young pulsars, whereas for old pulsars the best fit gives $L_{X,psr/pwne}\propto \dot{P}/P^{3.5}$,
the latter being
close to, and statistically indistinguishable from,  a constant
X-ray emission efficiency $\eta$;
2) young pulsars are more efficient X-ray emitters than old pulsars, and
have a different dependency for $L_X$ on $P$ and $\dot{P}$.

We also confirm the findings by \citet{arons96,abdo10} that
1) the \gray\ luminosity is well described
by $L_\gamma \propto \sqrt{\dot{E}_{rot}}$; 
2) younger pulsars are less efficient in producing \gray s.
The latter trend is, therefore, the reverse of the X-ray luminosity
(figure~\ref{fig:eta}).

The question is what these findings reveal about pulsar X-ray emission 
mechanisms.
First of all, the connection between the
X-ray luminosities of pulsars and their PWNe may be surprising,
given that the luminosities of the PWNe are, in general,
 affected by both the pulsar
wind properties and the environment of the pulsar. In particular for 
young PWNe the
luminosity is possibly 
affected by its interaction with a surrounding 
supernova remnant (SNR).
The reverse shock of the SNR will during a certain phase of the SNR evolution 
compress the PWN, 
which naively may be assumed to result in a brightnening of the PWN. However,
a recent study by \citet{gelfand09} showed the behavior of the
PWN luminosity to be more complex.
Their study indicates that the compression by the reverse
shock leads to a brightening of the radio luminosity,
but at the same time to an almost total 
quenching of the X-ray luminosity for a brief period 
(for the
specific model they calculated this happened between 18-30 kyr).
Apart from this brief phase, the X-ray luminosity traces the spin-down
luminosity surprisingly well, with an expected fluctuation of about 
0.3 decades in $\eta$.

Given that apparently the X-ray luminosities of the pulsars and their PWNe
seem to trace each other, and have similar dependencies on the
period and period derivative, one cannot easily 
use the  best-fit functions for young pulsars 
to derive a correlation
of X-ray luminosity with some well known physical pulsar property, such
as $\dot{E}_{rot} \propto \dot{P}/P^3$, $B_p \propto \sqrt{P \dot{P}}$, 
$B_{LC}\propto \dot{P}^{1/2}/P^{5/2}$ 
(the magnetic field
at the light cylinder),
$n_{GJ}\propto \sqrt{\dot{P}/P}$,  or $\dot{N}_{GJ} \propto \sqrt{\dot{P}/P^2} \propto \sqrt{\dot{E}_{rot}}$. 
For young pulsars the X-ray emission is poorly fit with a dependency
on $\dot{E}_{rot}^b$ (Table~\ref{tab:fits}).
This suggests that the X-ray luminosity from pulsars and PWNe may require
a more complex model than the \gray\ luminosity, with its 
$L_\gamma \propto \sqrt{\dot{E}_{rot}}$ scaling. The
best-fit formula does, however, explain the strong dependency
on $\tau_c$ for young pulsars, since
$L_X \propto \dot{P}^3/P^6 \propto \dot{E}_{rot}/(P\tau_c^2)$.

Secondly, the similar behavior of the pulsars and PWN X-ray luminosities suggest there is a physical
connection between the two. It
is usually assumed that the pulsar X-ray emission originates in the magnetosphere, whereas the X-ray emission
from the nebulae comes from outside the termination shock. 
These are two distinct regions, which are
separated by a region that encompasses the pulsar light cylinder and the so-called wind-zone \citep{kirk09},
the region in which the pulsar wind is formed.

It is tempting to speculate that the connection between the X-ray emission 
from the pulsars and PWNe
may have something to do with the pair multiplicity. Models of pair creation in the magnetosphere indicate that the
pair multiplicity is a function of the characteristic age 
\citep[e.g.][figure~6]{hibschman01}.
This would mean that both the X-ray emission from the pulsar and from the PWNe are somehow proportional
to the multiplicity. This is not a completely satisfactory explanation, because the X-ray synchrotron depends
on the total energy contained by the pairs, not just by the total number of particles.

Another explanation for the
similar behavior of pulsar and PWN luminosities is
offered
by the striped wind theory 
\citep[][]{coroniti90,kirk02,kirk09}. 
According to this theory the alternating magnetic fields generated by 
obliquely rotating pulsars leads
to reconnecting magnetic fields in the wind zone. 
This transforms magnetic energy
into kinetic energy, thereby changing the pulsar wind from a high to
a low $\sigma$ outflow. According to \citet{kirk02} the zone 
in which this heating and acceleration occurs could be the location of 
pulsed X-ray emission.  Such a model makes it easier to explain why
there is a connection between the pulsar and PWN luminosities, 
as in both cases the energy is generated in the wind zone.
The higher pair multiplicity  of young pulsars may be of additional 
importance as there
are simply more particles available to be accelerated.

Finally, there is one other issue to consider, namely, 
why young pulsars do not seem to be
efficient in generating \gray s. 
The fact that it is the other way around for the X-ray emission, 
perhaps indicates that the pair multiplicity in young pulsars is so high that 
the electric fields in the magnetosphere are shorted out \citep{hibschman01}. 
It may also be worthwhile to investigate in more detail
what the role of the seed photons for inverse Compton upscattering
in young pulsars is. The Compton scattered photons (and in some
cases curvature radiation generated photons) are converted in pairs,
and form the basis of the pair multiplicity \citep{hibschman01}.
The soft seed photons originate from the neutron star surface, and
may the result of polar cap heating due to accelerated electron/positron
beams, or they are the result of the cooling of the neutron star.
The latter would induce a dependency on neutron star age, which could
explain why young pulsars behave differently than old pulsars.

We have only briefly mentioned the TeV detected PWNe. 
The TeV luminosities of PWNe do not
seem to correlate with spin-down luminosity  \citep{mattana09}. 
The probable reason is  that
the TeV emission from PWNe is not so much determined by the current 
energy production of the pulsar, but reflects the time integrated energy 
input from the pulsar wind, which in itself
scales with the initial spin-period $E(t=0)$. In X-rays the
lifetime of the X-ray synchrotron emitting electrons is short compared to the 
pulsar lifetime, whereas in TeV (and in the radio) this is not the case. 

We can therefore conclude that different parts of the electromagnetic spectrum 
inform us about different aspects of pulsars. The \gray\ emission and pulsed
 radio emission is likely related to what happens in the magnetosphere, 
whereas TeV and radio emission from the PWNe inform us about the time 
integrated properties of pulsars. Based on our statistical study we have 
added here the suggestion that the X-ray emission from  the pulsar, like
the X-ray emission from the PWNe, may inform us about what happens in one of
the least
understood regions surrounding the pulsar: the wind zone.

\acknowledgements
JV is supported by a Vidi grant from the Netherlands Science Foundation (NWO).
RY is supported by grant-in-aid from the Ministry of
Education, Culture, Sports, Science, and Technology (MEXT) of 
Japan, No. 19047004, No. 21740184, No. 21540259.
We have made use of the ATNF online pulsar catalogue: 
\url{http://www.atnf.csiro.au/research/pulsar/psrcat}.

\begin{sidewaystable}
\caption{Best-fit solutions}
\begin{tabular}{llcccccc}\hline\hline\noalign{\smallskip}
                       & &  \multicolumn{4}{c}{X-ray\footnote{Based on \citet{kargaltsev08}, but omitting millisecond pulsars
and non-resolved PWNe (\#38,40).}} 
& $\gamma$-ray\footnote{Based on \citet{abdo10}, but omitting
millisecond pulsars and pulsars without distance estimates.} \\
       
                        && \multicolumn{2}{c}{Young\footnote{$\tau_c \leq 1.7\times 10^4$~yr}} & 
 \multicolumn{2}{c}{Old\footnote{$\tau_c > 1.7\times 10^4$~yr}} & \\
 Fit formula & Parameter\footnote{All quoted errors are rms errors.} & PWNe &PSRs & PWNe & PSRs &\\
\noalign{\smallskip}\hline\noalign{\smallskip}
(1) $\log L - \log \dot{E}_{rot} = a$ & $a$ 
& $ -3.26 \pm 0.33$ & $ -3.61 \pm 0.23$ & $ -3.94\pm 0.18$ &  $ -4.22 \pm 0.15$ & $-1.22\pm 0.15$
\\\noalign{\medskip}

(2)  $\log L =  a + b \log \dot{E}_{rot}$ 
& $a$ & $-37.2 \pm 12.2$  &  $-26.4  \pm 11.1$   &  $-14.2 \pm 7.12$ & $-1.65 \pm 6.32$ & $16.5 \pm 3.6$\\
& $b$ & $1.91 \pm 0.33$   &  $1.61\pm 0.29$      & $1.29\pm 0.20$   &  $0.93 \pm 0.18$ & $0.51 \pm 0.10$\\
& $F/P$\footnote{$F$-value and associated probability based on a comparison with formula 1 in this table.}
  & $7.56/1.2\%$          & $4.22/5.6\%$         &$2.1/17\%$ & $0.17/69\%$ & $24.7/0.004\%$\\

\noalign{\medskip}

(3)  $\log L =  a + b \log P + c \log \dot{P}$      &     
 a & $ 68.9 \pm 5.2$ & $63.9\pm 4.8 $   & $43.1\pm 3.9$   & $39.27\pm 3.5 $  & $40.1 \pm 2.4$ \\
&b & $-6.18\pm 0.72$ & $-5.23\pm 0.66$  & $-4.31\pm 0.78$ & $-3.18 \pm 0.70$ & $-1.56 \pm 0.39$\\
&c & $3.29 \pm 0.41$  & $2.84 \pm 0.38$ & $1.11 \pm 0.28$ & $0.77\pm 0.25$ & $0.49 \pm 0.18$  \\
&$F/P$\footnote{Based on a comparison with a fit using formula (1) in this table.} 
   &  $8.60/0.015\%$ &  $11.90/0.07\%$    & $1.41/27\%$ &  $0.46/64\%$ & $11.88/ 0.02\%$\\
&$F/P$\footnote{Based on a comparison with a fit using formula (2) in this table.} 
&  $16.85/ 0.08\%$ & $15.89/0.1\%$ & $0.78/39\%$ & $0.76/40\%$ &$0.017/90\%$\\
\noalign{\smallskip}
Sample size  & &\multicolumn{2}{c}{19}  & \multicolumn{2}{c}{18} & 28      \\
\noalign{\smallskip}\hline
\end{tabular}
\label{tab:fits}
\end{sidewaystable}

\end{document}